\documentclass[onecolumn,aps]{revtex4}

\usepackage{graphicx,color}
\usepackage{dcolumn}
\usepackage{bm}

\definecolor{grey}{rgb}{0.502,0.502,0.502}
\definecolor{orange}{rgb}{1.,0.5,0.}
\definecolor{brown}{rgb}{0.55,0.27,0.08}
\definecolor{dgl}{rgb}{0.,0.3922,0.}

\begin{document}

\title{Spontaneous knotting and unknotting of flexible linear polymers:\\equilibrium and kinetic aspects}

\author{L. Tubiana$^1$, A. Rosa$^2$, F. Fragiacomo$^3$ and C. Micheletti$^2$}
\affiliation{
(1) Department of Theoretical Physics, Jo\v zef Stefan Institute,\\ SI-1000 Ljubljana (Slovenia)\\
(2) SISSA - Scuola Internazionale Superiore di Studi Avanzati,\\ Via Bonomea
265, 34136 Trieste (Italy)\\
(3) Dipartimento di Fisica, Universit\`a degli Studi di Milano,\\
Via Celoria 16, 20133 Milano (Italy).
}

\date{\today}

\begin{abstract}
We report on a computational study of the statics and dynamics of long
flexible linear polymers that spontaneously knot and unknot.
Specifically, the equilibrium self-entanglement properties, such as the
knotting probability, knot length and position, are investigated with
extensive Monte Carlo sampling of chains of up to 15,000 beads. Tens of
such equilibrated chains of up to $\sim 4,000$ beads are next used as
starting points for Langevin dynamics simulations. The complex interplay
of chain dynamics and self-knotting is addressed by monitoring the time
evolution of various metric and entanglement properties. In particular,
the extensive duration of the simulations allows for observing the
spontaneous formation and disappearance of prime and composite physical knots in
linear chains. Notably, a sizeable fraction of self-knotting and
unknotting events is found to involve regions that are far away from the
chain termini. To the best of our knowledge this represents the first
instance where spontaneous changes in knotting for linear homopolymers are
systematically characterized using unbiased dynamics
simulations.
\end{abstract}

\maketitle

\section{Introduction}\label{sec:intro}
Polymer entanglement has been long studied for its impact on the statics
and dynamics of dense systems, of which artificial polymer melts and
tightly-packed biofilaments represent two notable instances.  One
particular form of entanglement is given by knots, which are ubiquitous
in long flexible chains~\cite{JPA89_Sumners_Whittington} and are known 
to affect their physical and functional
properties~\cite{Saitta_et_al:1999:Nature,Arai:1999:Nature,katritch1996knots,Orlandini:JPA:2008,Crisona:1999:J-Mol-Biol:10369759,RevModPhys.79.611,Marenduzzo:2010:J-Phys-Condens-Matter:21399272,PhysRevE.79.021806,PhysRevLett.99.217801,doi:10.1021/ma201827f}.

These aspects have so far been mainly studied for chains that form rings
by circularisation. In such case, in fact, the topological state of the
chain can be rigorously defined and can be analyzed with established
mathematical procedures~\cite{RevModPhys.79.611,meluzzi2010biophysics,Micheletti20111,grosberg2009}. 

In contrast to the case of circular chains, proper
  mathematical knots can not be defined in linear chains with free ends.
  Yet, we know by experience that localised, ``physical'' knots do appear in
  sufficiently-long open chains and can be long-lived. Indeed, the
  relatively-unexplored topic of physical knots is gaining increasing
attention because of its relevance in nanotechnological contexts,
especially regarding the implications for polymers mechanical
resistance, rheology and pore translocation capabilities.
In fact, recent related studies have addressed the problem of how a designed
tightly-knotted linear chain
disentangles~\cite{doi:10.1021/ma035100e,Gerland:NanoLett:2008,PhysRevLett.86.1414,MetzlerEPL2006},
how it responds to
stretching~\cite{Farago:2002:EPL,bao2003behavior,doi:10.1021_Huang,PhysRevE.81.041806,Kirmizialtin:2008:J-Chem-Phys:18331111,matthews2010effect,Matthews:2012:ACS-Macro-Lett:23378936},
and how knotted proteins~\cite{huang:121107} and knotted nucleic
acids~\cite{PhysRevLett.109.118301} translocate through a solid-state
nanopore.

Arguably, the simplest system where the influence of knotting on the
properties of linear polymers can be investigated is represented by long
flexible linear chains in equilibrium.
Such reference system has recently been studied to establish the equilibrium knotting
probability~\cite{Millett:2005:Macromol} and knot size of flexible
chains of beads consisting of up to 1000
monomers~\cite{virnau_et_al_knots_2005}.  However, to the best of our
knowledge, no study has yet focused on the kinetics of knotting and
unknotting in such system, nor in other homopolymeric systems.

Here, building on these previous studies, we extend the characterization of
the entanglement of unconstrained linear chains in two
directions. Specifically, we push the statics profiling to linear chains
composed of up to 15000 beads and, especially, we address the role of
chain dynamics in the knotting of flexible homopolymers. 

\begin{figure}
\includegraphics[width=4in]{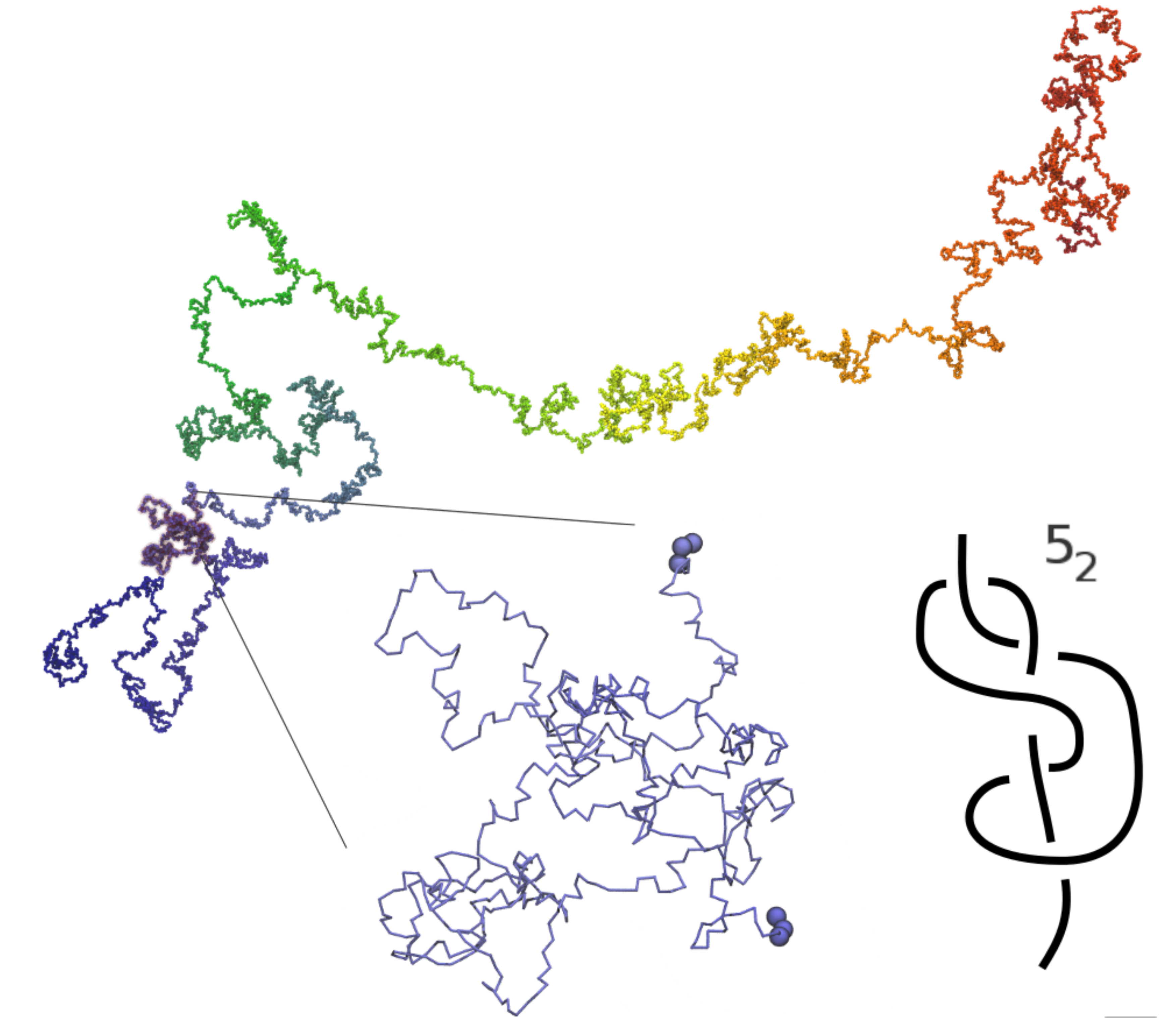}
\caption{ Typical configuration of a knotted chain composed by $N=8,192$
  beads sampled with the Monte Carlo scheme. The chain
    backbone is colored with a rainbow scheme along the contour. The
    expanded region is the knotted portion of the chain, which
    corresponds to a $5_2$ knot.  A simplified, schematic diagram of the
    $5_2$ knot is shown at the bottom on the right. For clarity, the
    chain backbone in the expanded view is shown with a thin line.  To
    represent the chain excluded volume, a few chain beads of the
    expanded region are explicitly represented.}
\label{fig:configuration}
\end{figure}

\section{Methods}\label{sec:modmeth}

The statics and dynamics of fully-flexible self-avoiding chains of beads
were characterized with stochastic numerical techniques. For
computational efficiency, Monte Carlo (MC) simulations were used for the
equilibrium properties, while kinetic ones were addressed with a
Langevin molecular dynamics (MD) scheme.
The two simulation techniques were applied to different, but physically equivalent, models of flexible self-avoiding chains were used, as detailed below.
 simulation techniques were applied to different

\subsection{Equilibrium: Monte-Carlo simulations.}\label{sec:MC}

The equilibrium properties were computed for self-avoiding
freely-jointed chains consisting of up to $N=15000$ spherical
beads. The beads diameter, $\sigma$, is taken as the unit of length.
 
Starting from a straight chain configuration, the conformational space
was explored using a standard MC scheme employing unrestricted
crankshaft and pivot moves \cite{Micheletti20111}.  A trial
configuration generated by either type of move was rejected if it
violated self-avoidance due to the presence of overlapping beads, and
accepted otherwise. All self-avoiding configurations were therefore
sampled with the same statistical weight.
A sample configuration of $N=8192$ beads is shown in Fig.~\ref{fig:configuration}.

For each chain length, we gathered $\sim10^5-10^6$ uncorrelated
configurations generated by the MC procedure, and used them to calculate
the expectation values of various observables, such as the fraction of
knotted chains, the incidence of various knot types and their contour
length, see section \ref{sec:observables}.

\subsection{Kinetics: Molecular Dynamics simulations.}\label{sec:MD}

Because of the fixed-bond length constraint, the freely-jointed chain
model is not well-suited for efficient MD simulations.  To this purpose
we therefore resorted to the flexible chain-of-beads model of Kremer and
Grest \cite{KremerGrestJCP1990}.  

In the following, the position in space of the center of the $i$th bead
is indicated with $\vec{r}_i$ while the distance vector of beads $i$ and
$j$ is denoted as $\vec{d}_{i, j} = \vec{r}_j - \vec{r}_i$ and its norm
simply as $d_{i, j}$.  The model Hamiltonian is:

\begin{equation}\label{eq:energy}
{\cal H} = \sum_{i=1}^{N-1} \left[ U_{\rm FENE}(i, i+1) + \sum_{j=i+1}^N U_{\rm LJ}(i,j) \right]\ ,
\end{equation}
where $i$ and $j$ run over the beads indices, $U_{LJ}$ enforces the excluded volume interaction between distinct beads (including consecutive ones), and $ U_{\rm FENE}$ enforces chain connectivity.
The complete expressions for the two terms are:
\begin{equation}\label{eq:fenepot}
  U_{\mathrm{\scriptscriptstyle{FENE}}}(i,i+1) = \left\{
\begin{array}{l}
- \frac{k}{2} \, R^2_0 \, \ln \left[ 1 - \left( \frac{d_{i,i+1}} {R_0} \right)^2 \right], \, d_{i,i+1} \leq R_0\\
0, \, \scriptstyle{d_{i,i+1} > R_0}
\end{array}
\right.\\
\end{equation}
\begin{equation}
U_{\mathrm{\scriptscriptstyle{LJ}}}(i,j) = \left\{
\begin{array}{l}
4 \epsilon \left[ \left( \frac{\sigma}{d_{i,j}} \right)^{12} - \left( \frac{\sigma}{d_{i,j}} \right)^6 + \frac{1}{4} \right], \quad\scriptstyle{d_{i,j} \leq \sigma 2^{1/6}}\\
0, \quad \scriptstyle{d_{i, j} > \sigma 2^{1/6}}
\end{array}
\right . 
\end{equation}

\noindent where $\sigma$ is the nominal bead diameter (which is the unit
length), $R_0=1.5 \sigma$, $k=30.0 \epsilon / \sigma^2$ and $\epsilon = 1.0 \, k_B\, T$,
where $k_B \, T$ is the thermal energy \cite{KremerGrestJCP1990}.
No hydrodynamic treatment is considered in the model.

The kinetics of chains of length $N$ up to $4096$ beads was studied
using fixed-volume and constant-temperature MD simulations with implicit
solvent.  The
dynamics was integrated with the LAMMPS engine \cite{lammps} with
Langevin thermostat. Periodic
boundary conditions were applied, with the simulation box chosen large
enough in order to avoid chain self-interactions across the
boundaries. The elementary integration time step was $\Delta t = 0.012\tau_{MD}$, where $\tau_{MD}=\sigma(m/\epsilon)^{1/2}$
is the Lennard-Jones time, $m$ is the bead mass (set equal to the LAMMPS
default value), and the friction coefficient, $\gamma$, corresponds to
$\gamma/m = 0.5 \tau_{MD}^{-1}$ \cite{KremerGrestJCP1990}.

\subsection{Observables}
\label{sec:observables}

{\bf Autocorrelation time.}  For an overall characterization of the
kinetics of the flexible, self-avoiding chains, we considered the time
autocorrelation function, $\phi(t)$, of the end-to-end distance vector,
$\vec{R}_{ee} = \vec{r}_N - {\vec r}_1$:

\begin{equation}
\phi(t) = \frac{ \langle \vec{R}_{ee}(t) \cdot \vec{R}_{ee}(0) \rangle }{ \langle \vec{R}_{ee}^2 \rangle }
\end{equation}

\noindent where $\langle \, \rangle$ denotes the average over simulation
time for 10 independent trajectories.
According to Rouse theory \cite{Doi&Edwards:1986}, the decay of $\phi(t)$ should be
described by a sum of exponentials with the slowest effective Rouse
decay time, $\tau_R$ scaling as $N^{1+2\nu}$. For chains that do not
experience self-avoidance the metric exponent $\nu$ is equal to 0.5, and
hence $\tau_R \propto N ^2$, while for self-avoiding chains $\nu \simeq
0.6$ and hence $\tau_R \propto N^{2.2}$.

To calculate the effective decay time of $\phi(t)$ in our MD simulations,
we first captured the expected theoretical behavior by fitting it
with a sum of two exponentials, $f(t)=a_0\,e^{-t/\tau_1} + (1-a_0)\, e^{-t/\tau_2}$, and then obtained $\tau_R$ by integrating the fitting
function, $\tau_R = \int_0^{\infty} f(t) dt = a_0 \tau_1 + (1-a_0)
\tau_2$.

{\bf Physical knots in linear chains.} The degree of entanglement of equilibrated
chains was characterized by computing the chain knotting probability,
that is the percentage of MC sampled chains that are knotted.
To compute this and other observables 
it is necessary to suitably extend the standard notion of knottedness. 
The latter is, in fact, rigorously
defined only for chains that are closed (or with suitably constrained
termini) as their topological state cannot be altered by continuous,
non-singular chain deformations respecting chain connectivity.

Linear chains can be assigned to a definite, or dominant, knotted
topology by bridging the two termini with an auxiliary arc, so to obtain
a closed chain for which the topological state is mathematically well-defined.
Several such closing procedure were previously introduced
\cite{0305-4470-25-24-010,Millett:2005:Macromol,virnau_et_al_knots_2005,PTPS.191.192.tubiana}.
Here we adopted the ``minimally-interfering'' closure scheme that was
recently introduced by some of us \cite{PTPS.191.192.tubiana}. In this
method, which is numerically efficient, the auxiliary arc is constructed
so to minimize the potential interference due to spurious entanglement
of the auxiliary arc with the rest of the linear chain.

After closure into a ring, the chain topology was established by first
simplifying the ring geometry with topology-preserving moves
\cite{PhysRevLett.66.2211,Taylor:Nature:2000,Micheletti20111}, and
finally computing the Alexander determinants $\Delta(t)$ in $t=-1$ and
$t=-2$.

\begin{figure*}[ht!] 
\includegraphics[width=7.0in]{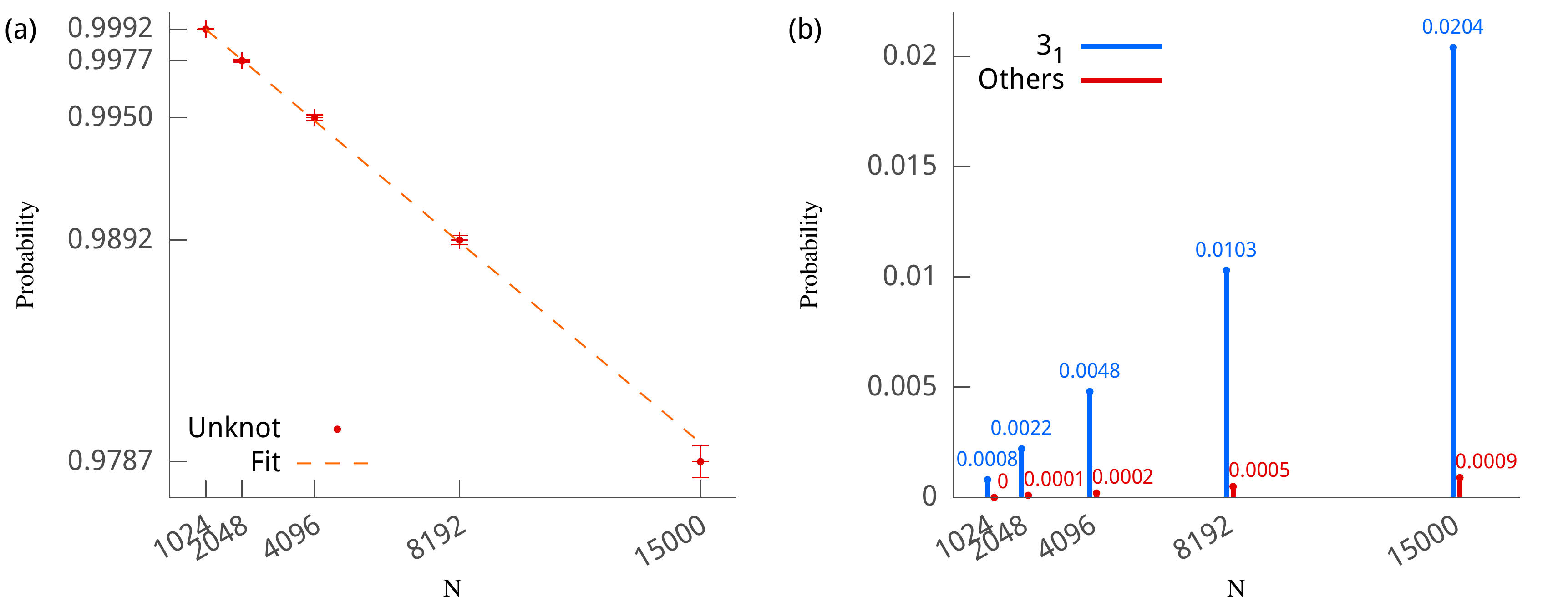}
\caption{ (a) The equilibrium fraction of unknotted chains,
  $P_{UN}$, is shown as a function of chain contour length, $N$. The
  solid line is the exponential best-fit, $A\,e^{-N/N_0}$ which yields
  $N_0 = (7.1 \pm 0.2)\times 10^5$.  (b) Probability of occurrence of trefoil and other knot types versus
  chain length.}
\label{fig:knot_prob}
\end{figure*}

{\bf Knot position and length.} The equilibrium and kinetic properties
of linear chains with non-trivial topology were further characterized by
establishing both the size and position of the knot(s) accommodated
along their contour. Specifically, we adopted the bottom-up knot search
strategy \cite{PhysRevE.61.5545,PTPS.191.192.tubiana}, which consists of
locating the shortest chain portion that, upon closure, has the same
topology as the whole chain. For a robust result, we required that the
knotted portion is strictly smaller than the whole chain and also
required that the arc formed by rest of the linear chain plus its
minimally interfering closure has (again upon closure) the unknotted
topology.

To minimize the computational cost of locating knots, the systematic
search of the shortest knotted arc was carried out by moving along the
chain contour in steps sizes of up to $N/100$. Here, the size of each
step is given by the number of bonds which can be rectified without
affecting the topology, according to the simplification scheme described
in Ref.~\cite{PTPS.191.192.tubiana}.

We expect the results to be largely independent of the
  specific knot search strategy because the amount of chain entanglement
  found {\em a posteriori} in our equilibrated open chains is limited
  (knotting probability not exceeding 3\% for the longest chains).  In
  this situation, in fact, different knot localization
  methods~\cite{PhysRevE.61.5545,virnau_et_al_knots_2005,Millett:2005:Macromol,Marcone2007size,doi:10.1021/ma980013l}
  usually yield consistent results (this may not hold for strong
  self-entanglement, see Ref.~\cite{PhysRevLett.107.tubiana}).


\section{Results and discussion}\label{sec:results}
\subsection{Equilibrium}\label{sec:equilibrium}

The MC scheme described in section \ref{sec:MC} was used to generate
equilibrated conformations of self-avoiding
freely-jointed chains of $N=$1024, 2048, 4096, 8192 and
15000 beads. At each chain length we sampled $10^5-10^6$ uncorrelated
conformations, which sufficed to gather $\sim$ 700--2000 independent
knotted chains.

The increase of chain self-entanglement with contour length is
illustrated in Fig.~\ref{fig:knot_prob}a which reports the fraction of
equilibrated configurations that are unknotted. This
unknotting probability, $P_{UN}$, is expected to decay 
exponentially with $N$
\cite{Michels08021986,JPA89_Sumners_Whittington,0305-4470-23-15-028,PhysRevLett.66.2211,koniaris:1991:JCP,shimamura:2000:JPA},
\begin{equation}
P_{UN}\simeq  e^{-N/N_0}\ .
\label{eqn:prob}
\end{equation}
The exponential fit of the data, shown by the dashed curve in
Fig.~\ref{fig:knot_prob}a, yields $N_0 = (7.1 \pm 0.2)\times 10^5$. This value is in good
agreement with the knotting probability previously reported by Virnau
{\em et al.} \cite{virnau_et_al_knots_2005} for flexible linear chains
of up to 1,000 beads (for this length it was found $P_{UN} \sim
99.91$\%, similarly to what found here). Furthermore, values with the
same order of magnitude, $N_0 \sim 8\times 10^5$, $N_0 \sim 2.1\times
10^5$ and $N_0 \sim 1.3\times 10^5$ were reported respectively for
closed self-avoiding chains of beads (as a limiting case of a rod-bead
model)~\cite{PhysRevLett.66.2211,koniaris:1991:JCP}, for rings on the
simple cubic lattice~\cite{van2002probability,1742-5468-2010-06-P06012}
and rings on the face-centered cubic lattice~\cite{0305-4470-23-15-028}.

\begin{figure}[htb!]
\includegraphics[width=4in]{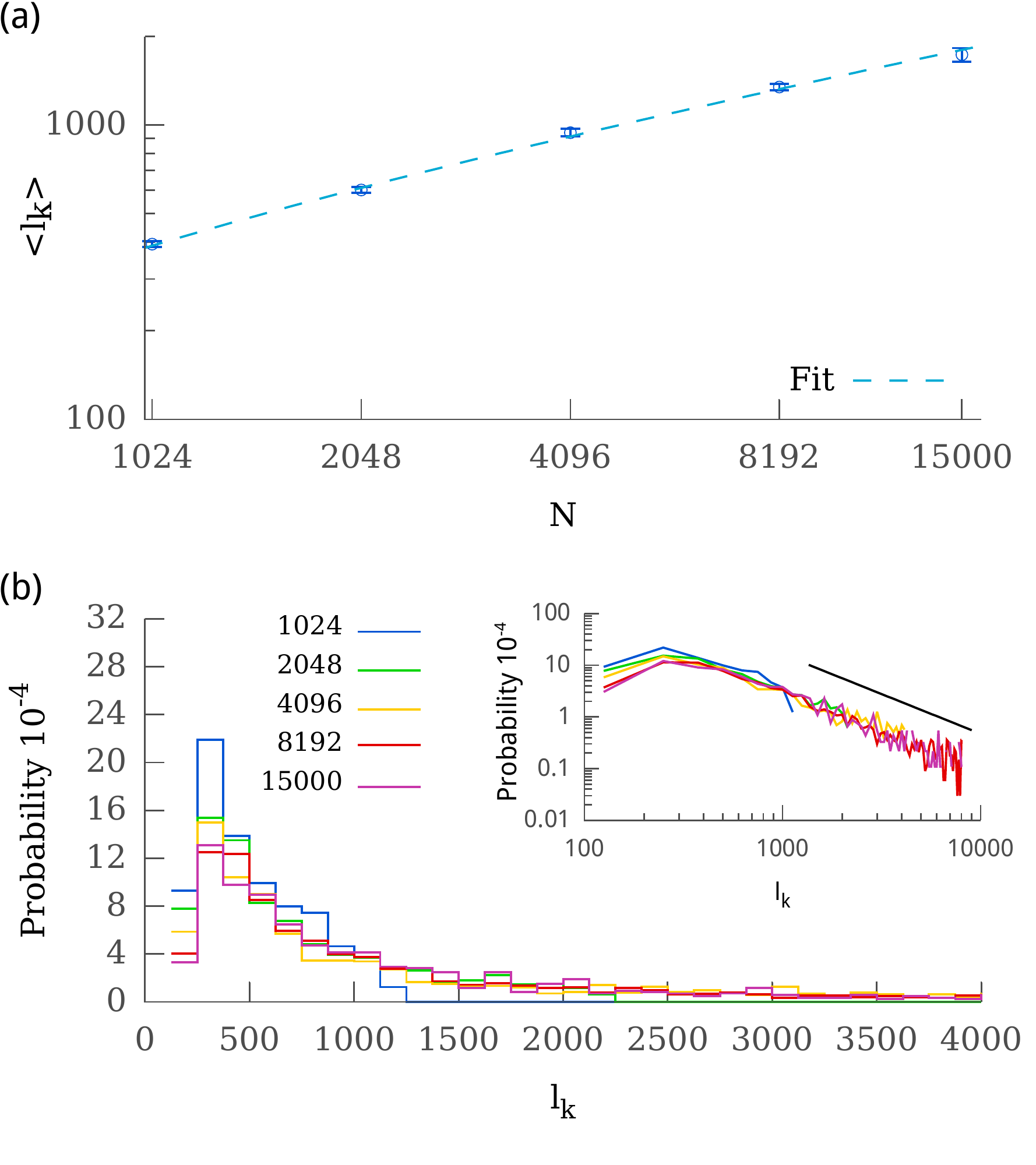}
\caption{ (a) Average knot length, $\langle l_k \rangle$, versus chain
  contour length, $N$.  The solid curve is the best-fit with the
  function $\langle l_k \rangle = a+ b\, N^\alpha$, with $\alpha = 0.44
  \pm 0.08$.  (b) Probability distribution of $l_k$ for various chain
  contour lengths. The distributions have an apparent
    linear trend in a log-log plot (see inset), which is suggestive of a
    power-law decay. The best linear fit of the log-log data for
    $l_k>1500$ and $N>1024$ yields the power law exponent, $-1.5 \pm
    0.1$. The associated power law is illustrated by the slope of the
    black guideline.}
\label{fig:klen}
\end{figure}

\begin{figure*}[tb!]
\includegraphics[width=7.0in]{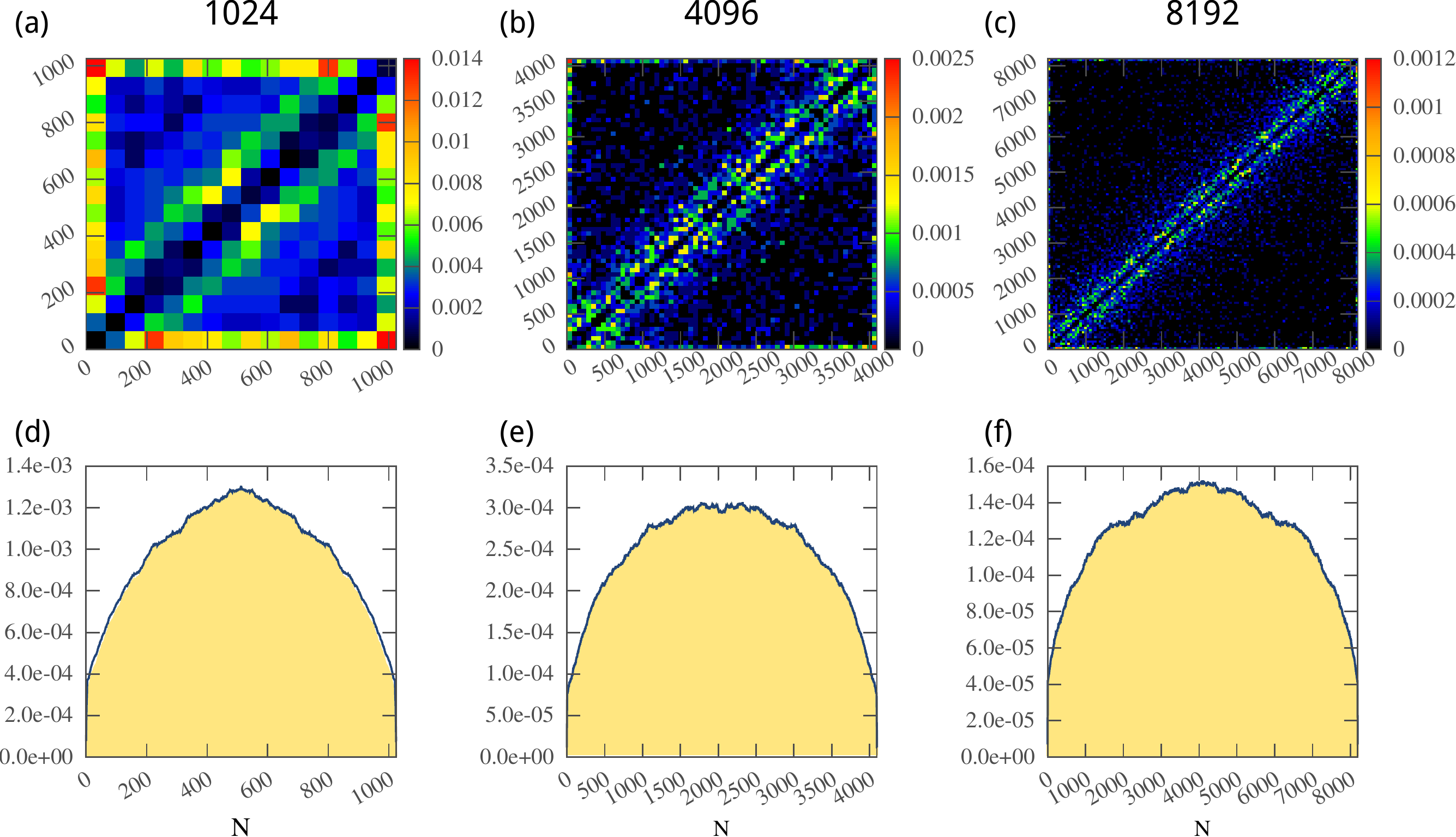}
\caption{ Equilibrium distribution of knots along
chains of $N=1024$, 4096 and 8192 beads.  After discretising the knotted
chains in segments of 64 beads we computed the (symmetrized) probability,
$p(i,j)$ that two given segments, $i$ and $j$ accommodate the ends of a
knot.  The resulting $p(i,j)$ matrices are shown as color-coded maps in
the top row panels a--c. We next calculated the probability that any given
chain bead falls within a knotted region. The resulting probability profiles for
various chain lengths are shown in panels d, e and f.  The
chain-reversal symmetry was used to improve the statistics of all the
plots.
}
\label{fig:densityplot}
\end{figure*}

The length-dependent increase of the chain knotting probability is
further associated to the appearance of knots of increasing
complexity. This fact, which is analogous to the case of closed
chains~\cite{PhysRevLett.66.2211,van2002probability,RevModPhys.79.611},
is illustrated in Fig.~\ref{fig:knot_prob}b.  It is seen that the
increasing incidence of the simplest knot type, the $3_1$ or trefoil
knot, is paralleled by the growth of more complex knot types. In
particular, at the two largest considered lengths, $N=8096$ and
$N=15000$ all knots of up to six crossings, with the exception of
$6_1$, are observed. In particular, we recorded the occurrence of
$3_1\#3_1$ composite knots, although their quantitative incidence is
minimal, ca. 20 instances out of the few thousands sampled knots.

The sizeable number of sampled configurations allows for computing the
expectation values and the probability distributions of various
observables, including some that were not considered in previous
investigations of knotted linear chains. A notable one is represented by
the average contour length of the knotted region, $\langle l_k \rangle$,
which can have important physical reverberations e.g. on the mechanical
resistance of a chain~\cite{Saitta_et_al:1999:Nature,Arai:1999:Nature}
or its capability to translocate through pores or
openings~\cite{huang:121107,PhysRevLett.109.118301}.

The dependence of $\langle l_k \rangle$ on the chain length, $N$, is
shown in the plot of Fig.~\ref{fig:klen}a.
It should be noted that, because trefoils are by far 
the dominant knot type for the considered range of $N$, the ensemble 
average $\langle l_k \rangle$ essentially reflects the average length of $3_1$ knots.

For closed self-avoiding chains it was previously shown that the average
length of trefoil knots (and other prime knots, too)
follows a power law, $\langle l_k \rangle \propto
N^{\alpha}$~\cite{JPA_Orlandini_1998,Marcone2007size,Mansfield:2010:1,PhysRevLett.107.tubiana}. Across
these studies, different values of $\alpha$ were reported but all of
them were strictly smaller than 1, which indicates a sublinear growth of
the average knot length with the chain contour length (weak knot
localization
\cite{Farago:2002:EPL,virnau_et_al_knots_2005,Millett:2005:Macromol}).

We observe that unconstrained linear chains display the same weak
localization property. In fact, the data for $\langle l_k \rangle$ are
well-interpolated by a power law with exponent $\alpha = 0.44 \pm 0.08$, see
Fig.~\ref{fig:klen}a. This exponent is compatible with the one estimated
by Farago {\it et al.}~\cite{Farago:2002:EPL} for mechanically-stretched
chains $\alpha=0.4$, though it differs from the one reported for linear
chains of up to $N=1,000$ beads, $\alpha
= 0.65$~\cite{virnau_et_al_knots_2005}. We note that the gap with this previously
  reported value can be bridged by considering only the data for $N \le 4096$, which yields $\alpha \sim 0.6$. This observation points out
  that current estimates of $\alpha$ may still be affected by
  appreciable finite size effects due to the fact that the accessible
  range of $N$ is still significantly smaller than $N_0$.

Valuable insight into the dependence of $\langle l_k \rangle$ on $N$
emerges by examining the probability distribution of $l_k$, which is
shown in Fig.~\ref{fig:klen}b.  First, it is seen that the location of
the peaks of the distributions show a very weak, if any, dependence on
$N$.  As a matter of fact, they all fall in the $100-300$ range.  Second,
the probability distributions extend appreciably beyond the peak value with a decay compatible with a power-law, see inset in Fig.~\ref{fig:klen}b.

Accordingly, knots with length exceeding by several times the most
probable (modal) value of $\langle l_k \rangle$ can occur with
non-negligible probability. These findings, which parallel the behaviour
of knotted chains subject to spatial confinement (see Fig. 4e in
Ref.~\cite{Micheletti:2012:SoftMatter}), indicate that
  the observed increase of $\langle l_k \rangle$ with $N$, results from
  the lack of a definite upper cut-off length of the distribution
  support. As a result, longer chains can accommodate a sizeable
population of knots with length that increasingly deviates from the most
probable one.  A similar effect was previously suggested for closed
rings in Ref.~\cite{Mansfield:2010:1}.

As a next step for characterizing the relationship between knot length
and chain length, we computed the probability that the two ends of the
knotted region fall in specific points of the chain.  The results are
shown in the density plot of Fig.~\ref{fig:densityplot}. It is seen that
the probability distribution is fairly uniform in the chain
interior. Apart from a localised enhancement very close to the two
termini (possibly due to effect of the auxiliary closing arc), the
probability density drops near the chain ends.  The width of this
depletion region at the chain edges is largely independent of $N$, and
about equal to 200.  The latter value about coincides with the typical
modal value of $l_k$, suggesting that the depletion arises because of
the impossibility to fully accommodate knots with the most probable
length near the chain ends.  

To further characterize the relation between chain length and knot
length we computed the probability that any given chain bead falls
within a knotted region. The probability profiles, which were calculated
considering only knotted chains, are shown in
Fig.~\ref{fig:densityplot}d--f for chains of length $N=$ 1024, 4096 and
8192, respectively. The appearance of the distributions is
center-symmetric and unimodal and the central region acquires a flatter
character for increasing $N$. The properties of the profiles are consistent
with the above-mentioned depletion effect near the chain edges.  

\subsection{Dynamics}

Following previous studies of knotted closed
chains~\cite{Orlandini:JPA:2008,Quake:PRL:1994,Sheng:PRE:1998,mansfield:2010:2}, 
we start the dynamics analysis by comparing the
characteristic timescales over which metric and topological properties
evolve. Next we address how the knotting properties
observed in equilibrium are linked to the kinetics of spontaneous
formation or untying of knots and their change in size and
position along the chain contour.\\

 The characterization of these kinetic properties is based on Langevin
 dynamics simulations for flexible, self-avoiding
 chains of $N = 1024$, 2048 and 4096 beads. As described
   in the Methods section, the chain model employed in the Langevin
   simulations is physically-equivalent, but not identical to the one
   used in Monte Carlo simulations. In particular, the constraints of
   fixed bond length and sharp excluded volume interactions are softened
   to be amenable to numerical dynamics simulations.

To optimally monitor the temporal persistence of knots and their motion
along the chain contour, we used a specific subset of MC-generated
chains as starting configurations.  Specifically, we focused on chains
which, in their center, accommodate a trefoil knot with length $l_k$ in
the $100-300$ range. The trefoil topology was picked because it is the
dominant one at the considered chain lengths, while the length range was
chosen because it straddles the most probable knot length for all values
of $N$, see Fig.~\ref{fig:klen}.  For each chain length we considered
about 30 different initial conformations.

\subsection{Metric and topological autocorrelation times}

\begin{figure}[h!]
\includegraphics[width=4in]{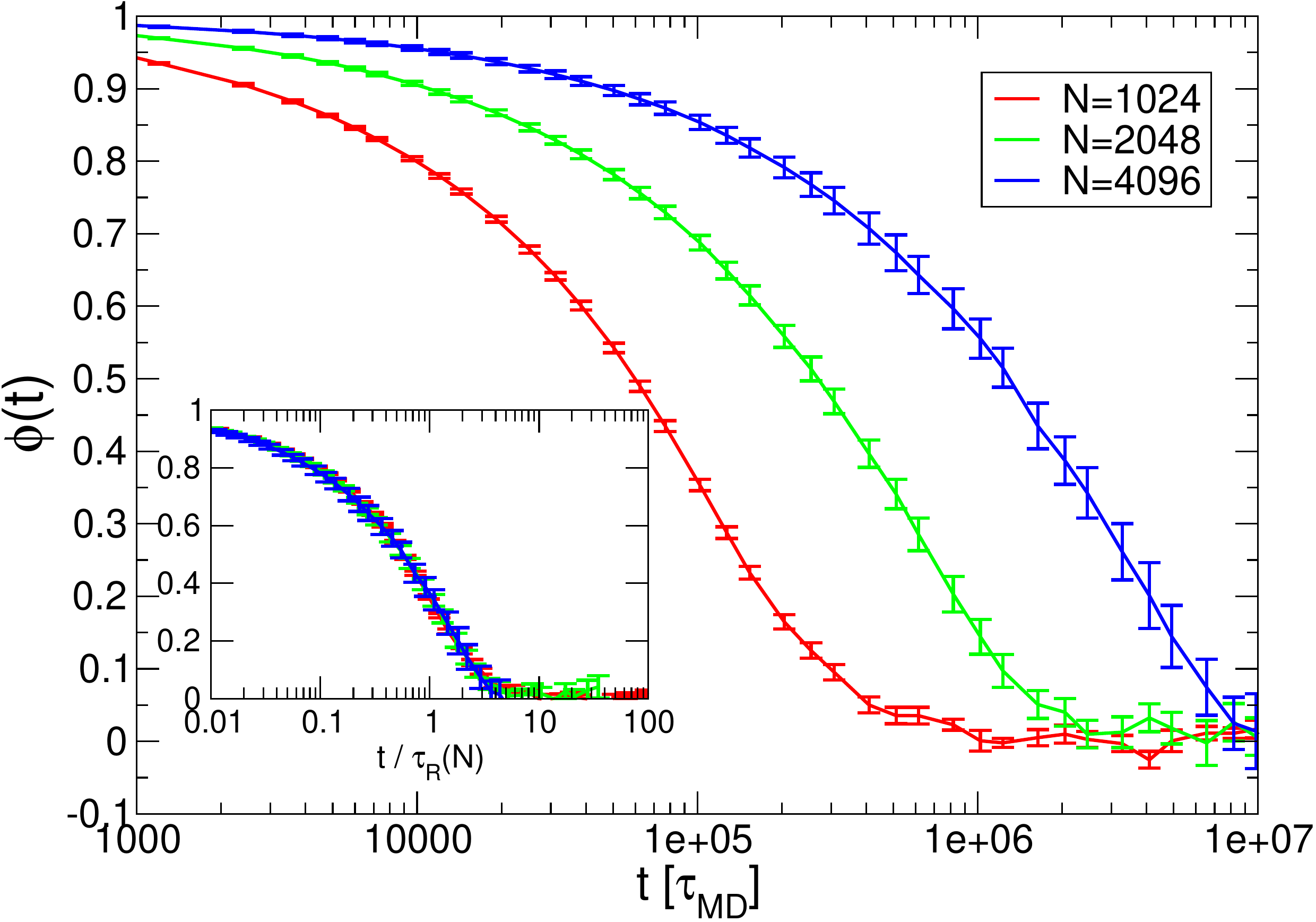}
\caption{Time autocorrelation function, $\phi(t)$ of the end-to-end
  distance vector. The data points are obtained by averaging over 10
  independent extensive Langevin dynamics trajectories. The bars indicate the
  standard error of the mean.  Inset: the good superposition of the
  rescaled autocorrelation functions, $\phi(t/\tau_R(N))$ confirms that
the decay of $\phi(t)$ is controlled by one dominant time scale
compatible with theoretical Rouse time, $\tau_R$.}
  \label{fig:etoe}
\end{figure}

As a term of reference we first computed the characteristic timescale
for chain metric properties. To do so we calculated the end-to-end distance
autocorrelation function, $\phi(t)$, over extensive MD simulations, see
Fig.~\ref{fig:etoe}. The characteristic decay time of $\phi(t)$,  the so-called Rouse time, $\tau_R$ is about
$1.04 \times 10^5 \, \tau_{MD}$,
$\simeq 4.7 \times 10^5 \, \tau_{MD}$ and
$\simeq 2.25 \times 10^6 \, \tau_{MD}$
for $N=1024$, $2048$ and $4096$, respectively. Bearing in mind the limitations of the
small set of considered chain lengths, the power-law best fit gives
$\tau_R \propto N^{2.25}$, in good accord with theoretical
arguments \cite{Doi&Edwards:1986}, see Methods.

\begin{figure}[h!] 
\includegraphics[width=4in]{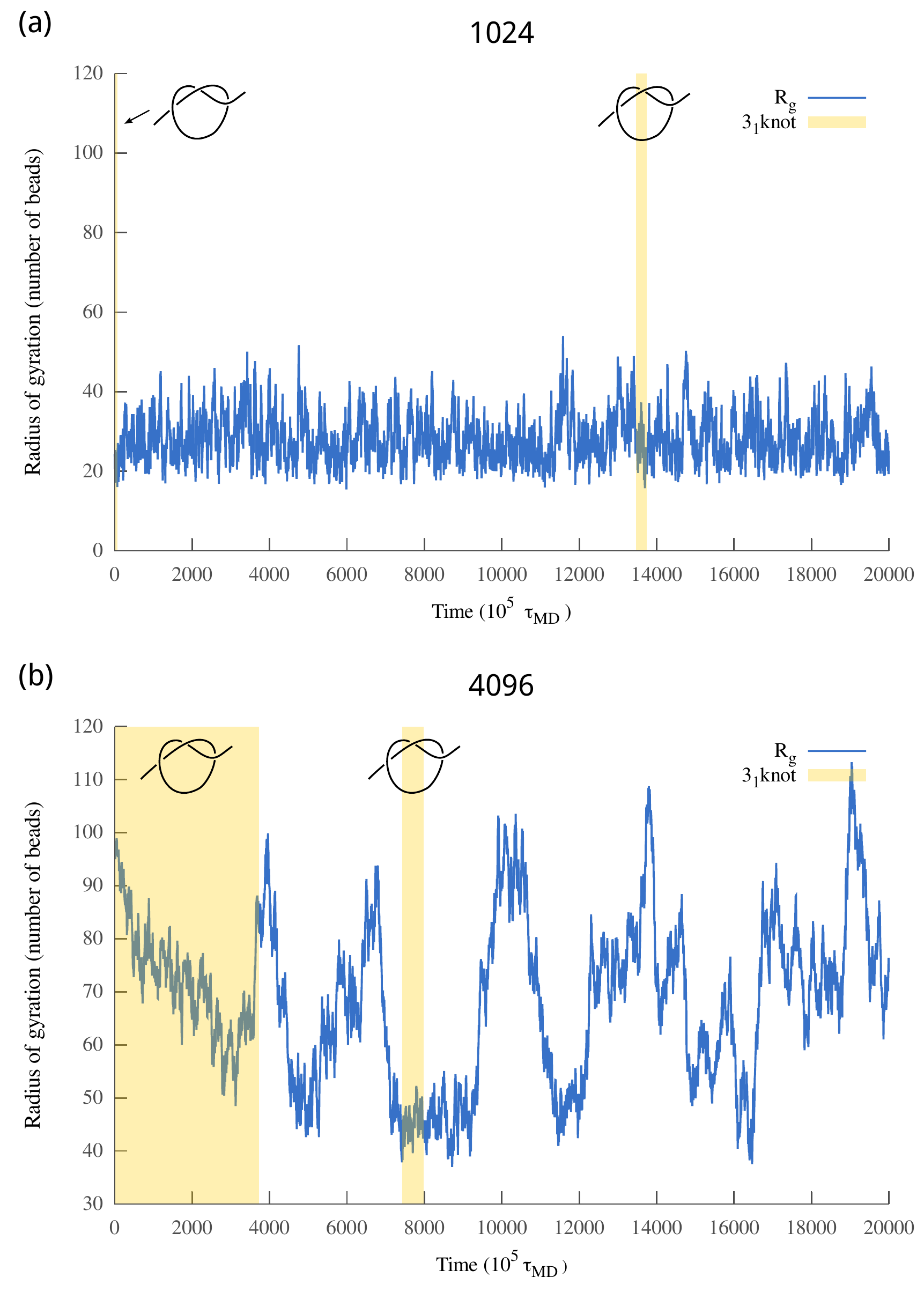}
\caption{Typical evolution of the metric and topological properties of
  two initially-knotted chains of (a) $N=1024$ and (b) $N=4096$
  beads. In both cases the initial configuration accommodated a trefoil
  knot at the chain center. The metric properties are captured by the
  chain radius of gyration (blue curve) while the time intervals where
  the chain is knotted are shown with a colored background. Spontaneous
  unknotting and knotting events are clearly visible. In both cases the
  spontaneously-formed knots have the dominant, $3_1$ topology.}
\label{fig:unknotting}
\end{figure}

The typical evolution of chain topology over time scales much longer
than $\tau_R$ is illustrated in Fig.~\ref{fig:unknotting}, which also
indicates with a colored overlay, whether knots are present or absent
during the dynamical evolution.

In both examples, it is seen that after the untying of the initial
knot the chain remains mostly unknotted for the rest of the simulation,
with the exception of transient self-knotting events.
The sparse occurrence of knots is consistent with the low
incidence of knots observed in equilibrium (which should coincide with
the time-averaged knotting probability in asymptotically-long dynamical
trajectories), see e.g. Fig.~\ref{fig:knot_prob}.

\begin{figure}[h!] 
\includegraphics[width=4in]{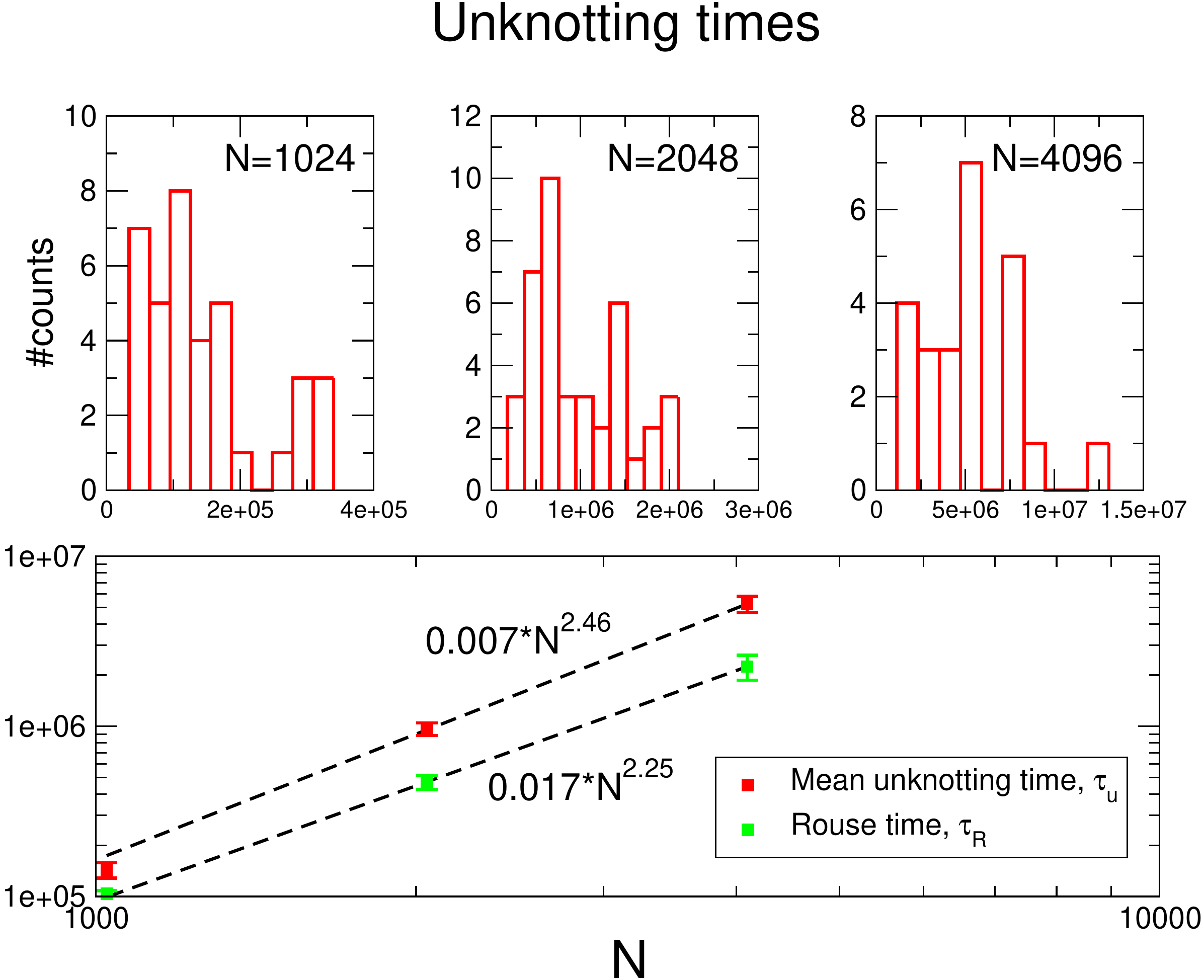}
\caption{ (Top) Distribution of unknotting times for chains of $N=1024$,
  2048 and 4096 beads.  (Bottom) Scatter plots of the mean unknotting
  (red symbols) and Rouse (green symbols) times versus $N$.  The dashed
  curves are the power-law best fits.  Times are given in units of
  $\tau_{MD}$.}
\label{fig:unknotting_times}
\end{figure}

The distribution of times required to spontaneously untie the trefoil
knot initially centered in the middle of the chain, is shown in the top
panels of Fig.~\ref{fig:unknotting_times} for various values of $N$. It
is seen that the broad range of unknotting times covered by the
distributions grows noticeably with $N$.  Indeed, the average unknotting time, $\tau_u$ takes on the
values $\simeq 1.43 \times 10^5 \, \tau_{MD}$, $\simeq 9.67 \times 10^5
\, \tau_{MD}$, $\simeq 52.5 \times 10^5 \, \tau_{MD}$, for $N=1024$,
2048 and 4096, respectively. As shown in the bottom panel of
Fig.~\ref{fig:unknotting_times} these values are not only larger than
the corresponding Rouse times, but their scaling, $\propto N^{2.46}$, is
also characterized by a larger exponent than $\tau_R$.

Although the uncertainty on the effective scaling exponent is sizeable,
15\%, due to the limited set of
considered chain lengths, the visual inspection of the trends of
$\tau_R$ and $\tau_u$ in Fig.~\ref{fig:unknotting_times} supports the
faster increase of $\tau_u$ over $\tau_R$.  Indeed, $\tau_u/\tau_R$
ranges from $\simeq 1.4$ for $N=1024$ to $2.3$ for $N=4096$.

Because this different growth is suggestive of a non-trivial interplay
of changes in chain geometry and the modes of chain unknotting, we
believe it would be most interesting to further address this specific
point in future studies by considering much longer chains. Further
motivations for such extensions are provided in the next sections in
connection with the mechanism presiding the spontaneous formation or
untying of knots and their motion along the chain contour.

\subsection{Spontaneous knotting and unknotting events}

To elucidate the mechanisms leading to spontaneous knotting and
unknotting events, we have monitored the knots position and length during
the simulations.

\begin{figure*}[t]
\includegraphics[width=7.0in]{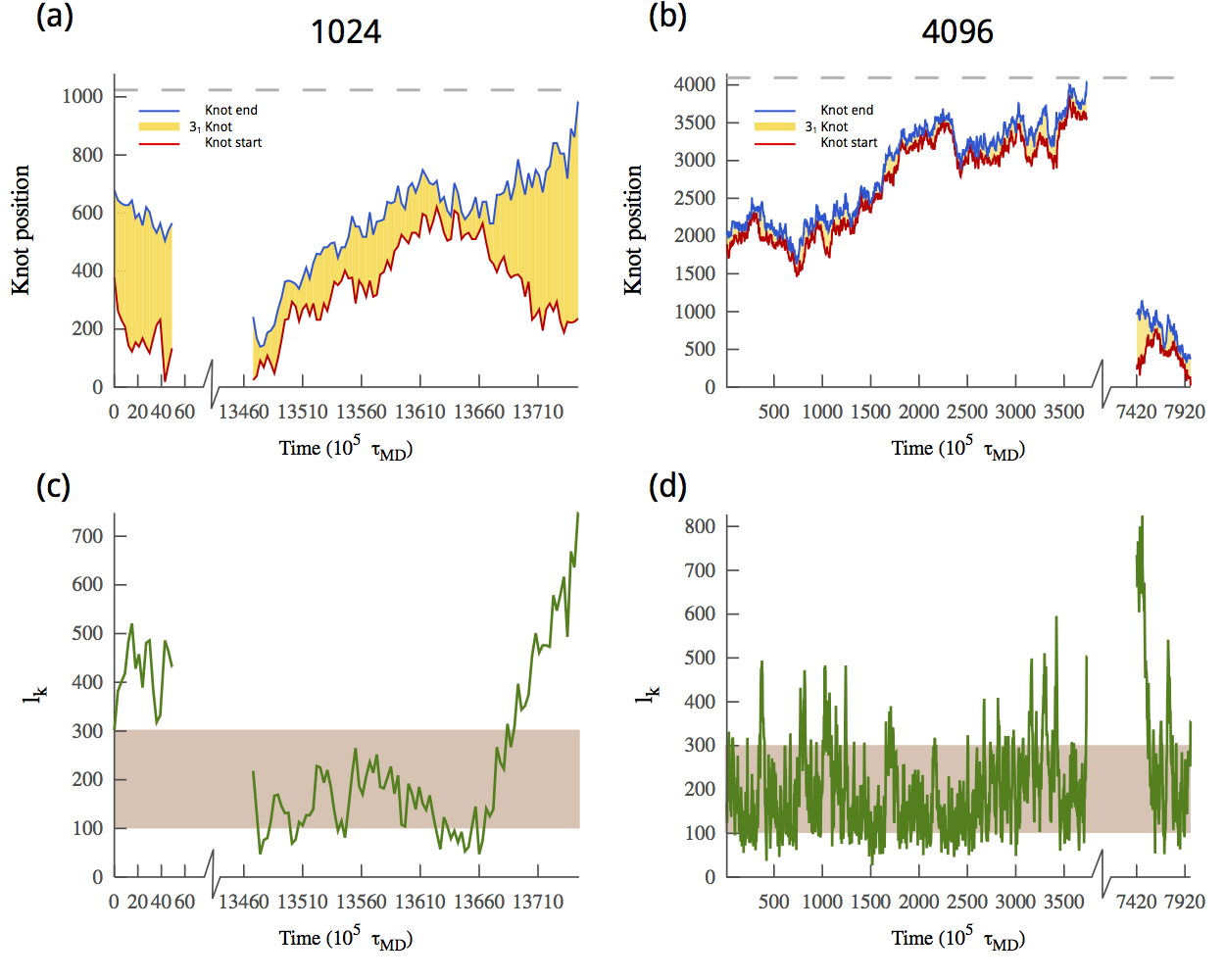}
\caption{Detailed characterization of the knotting kinetics for the two
  trajectories shown in Fig.~\ref{fig:unknotting}.  The upper panels
  illustrate the time-evolution, along the chain contour, of the two
  ends of the knot, marked in red and blue. The knotted region is
  highlighted in yellow. The evolution of the knot contour length, $l_k$,
  is shown in the bottom panels (c) and (d). The brown overlay
  highlights the most probable range of $l_k$, see
  Fig.~\ref{fig:klen}.}
\label{fig:knot_position_length}
\end{figure*}

\begin{figure}[t] 
\includegraphics[width=4.5in]{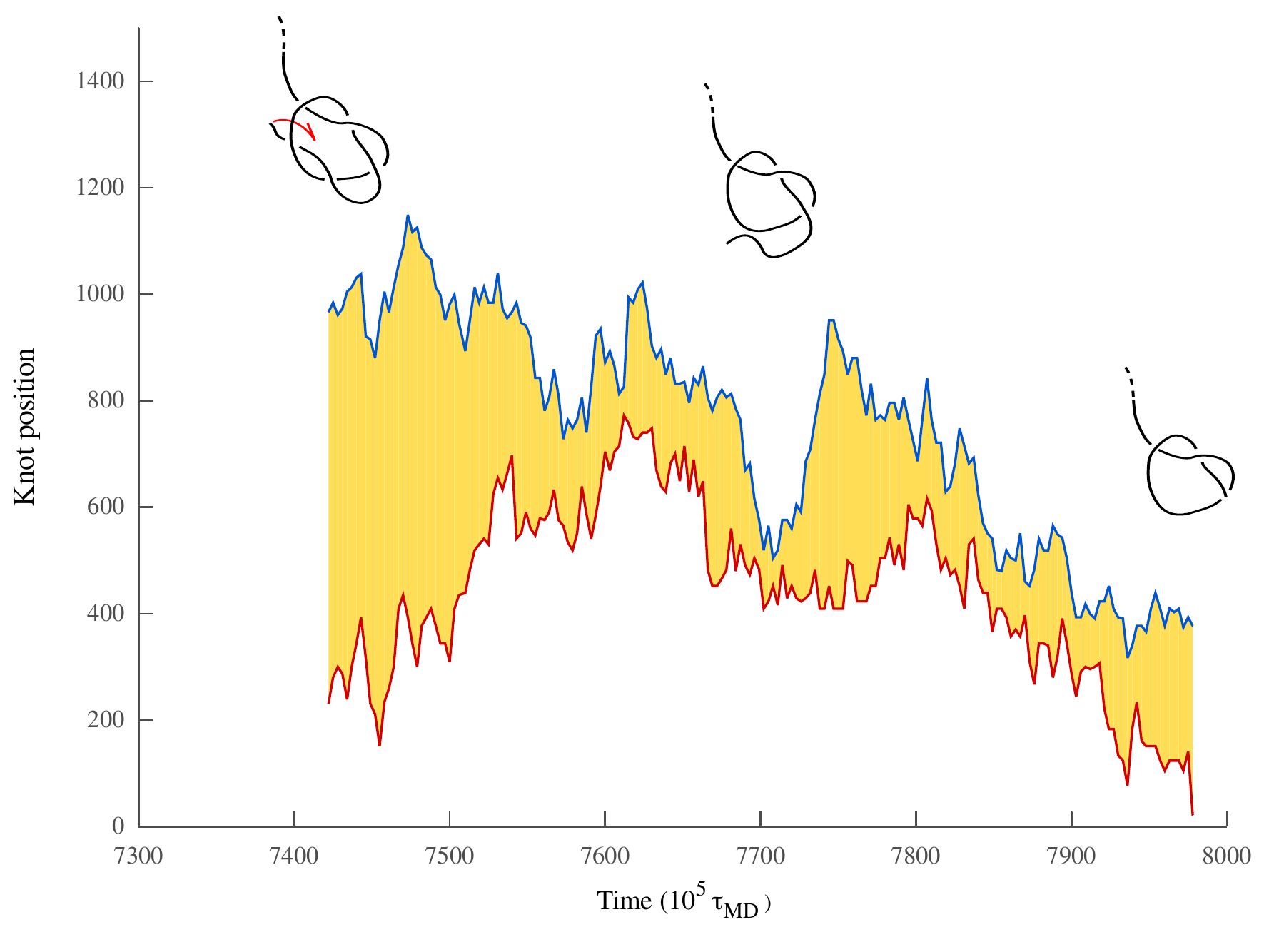}
\caption{Detailed characterization of the spontaneous knotting event
  shown in Figs.~\ref{fig:unknotting}b and
  ~\ref{fig:knot_position_length}b for a chain of $N=4096$ beads. The
  knot suddenly appears at a finite distance (ca 200 beads) from the
  nearest chain end. The mechanism of knot formation is shown
  schematically in the upper plot and involves a slipknot stage. Notice
  that the resulting trefoil is atypically large as it covers about one
  third of the chain contour.}
\label{fig:slipknotting}
\end{figure}
 
Fig.~\ref{fig:knot_position_length} presents these quantities for the
two trajectories previously illustrated in Fig.~\ref{fig:unknotting}.
It is seen that the knot moves away from its initial central location
with a stochastic motion along the chain contour. The motion of the two
knot ends, though not exactly concerted, is visibly correlated. In fact,
as illustrated in panels (c) and (d) of
Fig.~\ref{fig:knot_position_length}, the knot length, $l_k$, mostly
fluctuates in the range covered by the peak of the equilibrium $l_k$
distribution, see Fig.~\ref{fig:klen}.

By examining the traces in Fig.~\ref{fig:knot_position_length}(a), (b) one observes
that the first unknotting event occurs when one of the two knot ends
reaches one of the chain termini. Analogous considerations can be made
for the other changes of topology illustrated in the two panels. We
emphasize that this is the first time that
spontaneous knotting events are systematically observed in unbiased MD
simulations of a general homopolymer model.

Notice, in particular, that for the illustrated $N=1024$ case, a trefoil
knot is formed at one end, travels to the center of the chain where it
swells along the chain contour and finally unties by slipping off the
other chain end. The lifespan of this knot, which crosses the whole
chain, is larger than the time required for the initial knot (positioned
in the chain center) to untie. Throughout the collected simulations,
such long-lived knots are uncommon. In fact, the typical ``life-cycle''
of a knot is akin to the one shown for the longer chain, $N$=4096, in
panel (b). As it can be seen, the knot which is formed spontaneously at
one end, briefly diffuses towards the chain interior before slipping out
of the same end from where it originated.

While most of the knots spontaneously form or untie at the chain ends, a
small, but sizeable, fraction of knotting/unknotting events occur via a
mechanism that, borrowing the terminology introduced in protein-related
contexts~\cite{King:JMB:2007,Sulkowska:PNAS:2009}, can be termed as
slipknotting.

These knotting events occur when a hairpin-bent portion of the chain
enters a pre-formed loop and drags the entire terminal segment through
it (while the time-reversed procedure leads to unknotting). The
mechanism, which was recently reported and discussed in folding
simulations of knotted proteins~\cite{Sulkowska:PNAS:2009} is sketched
in Fig.~\ref{fig:slipknotting}. This figure also illustrates the
time-evolution of the two ends of a trefoil knot that forms via
slipknotting and dissolves by slipping out of the chain.  Notice that,
at variance with the more intuitive case of knots forming or
disappearing at the chain ends, slipknotting events do not necessarily
occur near the chain termini. As illustrated in
Fig.~\ref{fig:slipknotting} they, in fact, generally manifest by the
sudden appearance of a knotted region far away from the termini.

The relevant question that emerges from the results discussed above
regards the relative incidence of the two types of knotting/unknotting
events, namely those occurring at the chain termini or via slipknotting.
To this purpose, we gathered 90 extensive trajectories throughout the
considered range of $N$ and counted how many of their $134$ spontaneous
events of knot formation/disappearance occurred at a distance of at
least 100 beads from the chain termini. This threshold length was chosen
because it exceeds by ten times the typical contour distance traveled by
the knot ends in two consecutive snapshots of the recorded
trajectories. The criterion therefore provides a conservative counting
of the number of slipknotting events. It is found that 11 of the
topology changes, corresponding to $\sim 10\%$ of all knotting or
unknotting events, involve slipknotting.

This clarifies that slipknotting, though 
not representing the dominant
knotting mode for the considered flexible linear chains still accounts
for a sizeable fraction of their spontaneous topology changes.

Furthermore, because the two mechanisms differ for their local/non-local
character it may be envisaged that their relative incidence could be
significantly affected by the length of the chain. As a matter of fact,
none of the observed slipknotting events took place in the shortest
chains, $N=1024$, but occurred exclusively for $N=2048$ and $N=4096$.
This provides a further element of interest to address in future
investigations of knotted chains. In particular, one may anticipate that
the balance of the two mechanisms could have important reverberations
regarding the dependence of the unknotting time on $N$, too.

\begin{figure*}[floatfix] 
\includegraphics[width=7in]{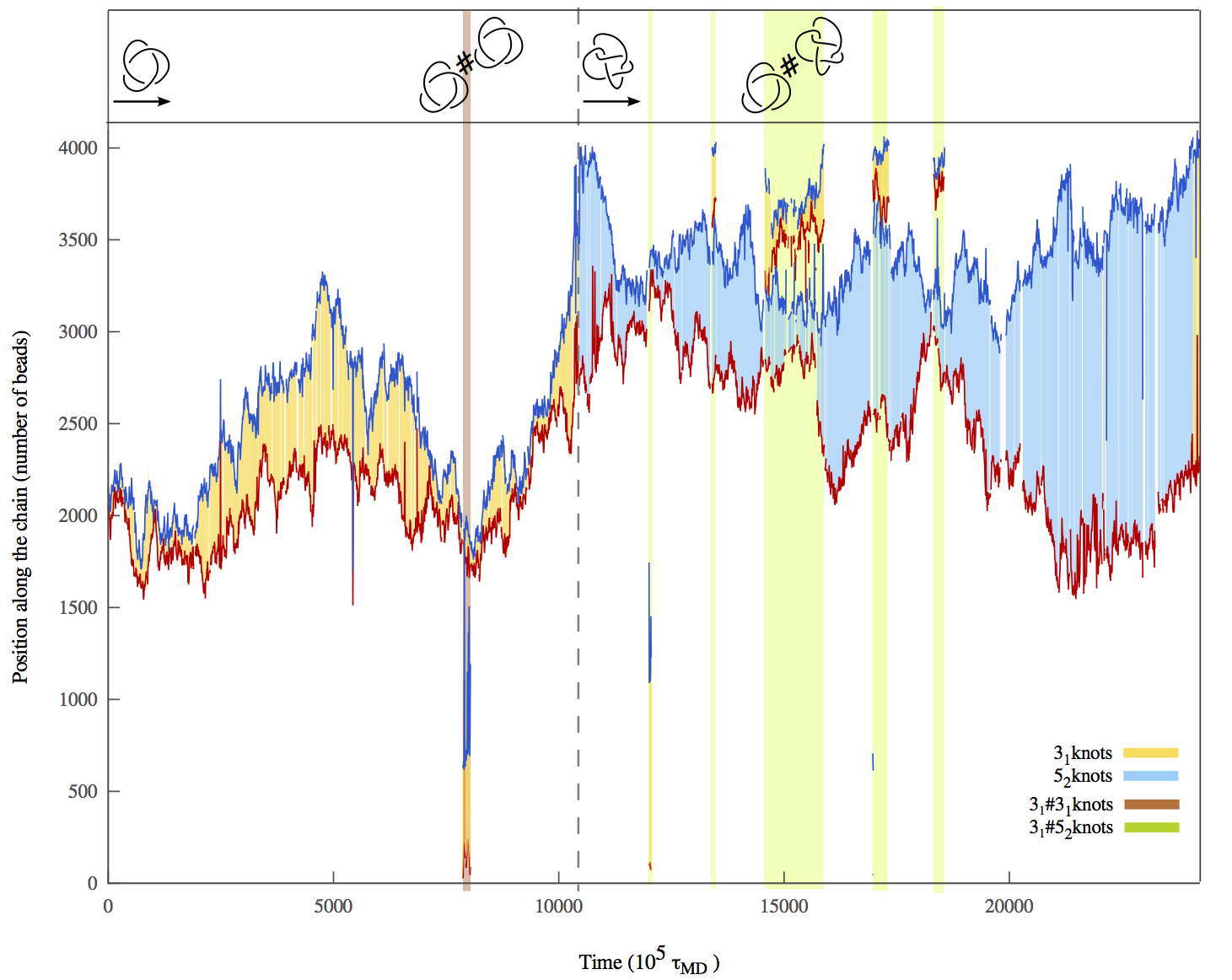}
\caption{Dynamical evolution of a chain of $N=4096$ beads exhibiting
  several transitions of knots topologies. In particular, one observes
  both changes in knot topology ($3_1 \to 5_2$) as well as the
  formation of composite knots. In the latter case, the separate prime
  components are well identifiable as shown by the colored overlays.
  The various encountered knotted topologies are illustrated at the top
  with schematic conventional knot diagrams. Notice that the
  conventional ``ring'' diagrams are used while the considered chain is
  actually linear.}
\label{fig:knotting_events}
\end{figure*}

\subsection{Formation of composite knots}

We conclude the analysis of the system dynamics by discussing notable,
though rare, instances where chain topology changes by the incremental
addition (and subsequent removal) of entanglement to an already-knotted
chain.

The trajectory shown in Fig.~\ref{fig:knotting_events} provides a
remarkable illustration of this point. In fact, one first observes
that the initial $3_1$ knot evolves transiently into a $3_1\#3_1$
composite knot due to the temporary formation of an additional trefoil
knot at one of the termini. At a later time, after reverting to the
$3_1$ topology, the knot becomes a $5_2$ knot. Next, a further
trefoil-knotting event occurs through slipknotting and the chain
acquires a $5_2\#3_1$ topology. Finally, after an intermittent
formation/disruption of the slipknotted trefoil (arguably due to a
persistent chain geometry) the chain goes back to the $5_2$ topology,
then the $3_1$ one and finally it unties itself.

One aspect that is highlighted in Fig.~\ref{fig:knotting_events} is that
the composite knots $3_1\#3_1$ and $3_1\#5_2$ present well separated
prime components. This fact is consistent with the
  observed dynamics of knot formation: because knots tend to originate
  at chain ends and their length grows sublinearly with chain contour length,
  one can expect that composite knots can form by incremental addition
  of individual prime components. Aside from these kinetic mechanisms,
  we note that across the limited number of composite knots sampled with
  the equilibrium MC simulations, no appreciable overlap was found for
  the individual prime components.

\section{Summary and Conclusions}\label{sec:concl}

We have presented a numerical study of the spontaneous occurrence of
knots in free long flexible chains of beads.  The study was carried out
at two levels. First we addressed the equilibrium entanglement, that is
the knotting probability, knot size and location along the chain, by Monte
Carlo sampling of chains with up to $N=$15000 beads. This represents a
more than tenfold extension in length over previous studies of
entanglement in flexible linear chains. Secondly, by using extensive
Langevin dynamics on chains with up to $N=4096$ beads, we provide the
first investigation of the connection between the equilibrium
entanglement properties and the free polymer dynamics.

For the equilibrium properties we find that characteristic length
controlling the $N$-dependent exponential decrease of unknotted chains
is $N_0 \sim 7.1\times10^5$ beads (implying a 2\% knotting probability
for the longest chains, $N=15000$). The knotting probability profile is
depleted near the ends of the chain over a region of width of about 200
beads. This boundary effect appears related to the most probable size of
knots, which falls in the 100-300 beads range independent of the chain
contour length. However, because the knot length probability
distribution decays slowly the average knot size does increase with $N$,
albeit sublinearly (weak knot localization).

Regarding the dynamical properties, over an extensive set of simulated
trajectories, we observed the occurrence of more than 100 spontaneous
knotting and unknotting events. These changes in topology typically
involve the chain ends, when a loop is threaded (or unthreaded) by one
of the chain termini. However, about 10\% of the knotting or unknotting
events are shown to form via a different mechanism which causes knots to
tie or untie away from the chain termini.

Because of the relative weight these two mechanisms could depend on the
chain contour lengths, we think that it would be most interesting to
address this point in future studies by extending the chain length. By
doing so it also ought to be possible to characterize the impact of
these mechanisms on the knot mean life time and its relationship with
the chain global relaxation time.

\section{Acknowledgements}\label{sec:ack}

We are indebted to Marco Di Stefano and Enzo Orlandini for valuable discussions. We acknowledge
financial support from the Italian Ministry of Education, grant PRIN
2010HXAW77.

\section{Notice}\label{sec:note}
This document is the unedited Author's version of a Submitted Work that was
subsequently accepted for publication in
Macromolecules, copyright \copyright American Chemical Society after peer review. To
access the final edited and published work see
http://dx.doi.org/10.1021/ma4002963 .
\bibliography{./knot_localization_open_chains_v2}

%

\end{document}